\documentclass[twocolumn,pre,twoside,groupedaddress,floatfix]{revtex4-1}

\usepackage{amsmath}
\usepackage{amssymb}
\usepackage{graphicx}
\usepackage{dcolumn}
\usepackage{bm}
\usepackage{color}
\usepackage{soul}

\def\dps{\displaystyle}

\def\d{\mathrm{d}}

\def\epsilon{\varepsilon}
\def\theta{\vartheta}
\def\rho{\varrho}

\def\set#1{\underline{#1}}
\def\vec#1{\mathbf{#1}}

\begin{document}


\title{Local pressure for confined systems}

\author{Paolo Malgaretti}
\email{malgaretti@is.mpg.de}
\author{Markus Bier}
\email{bier@is.mpg.de}
\affiliation
{
   Max Planck Institute for Intelligent Systems, 
   Heisenbergstr.\ 3,
   70569 Stuttgart,
   Germany, 
   and
   Institute for Theoretical Physics IV,
   University of Stuttgart,
   Pfaffenwaldring 57,
   70569 Stuttgart,
   Germany
}

\date{11 January 2018}

\begin{abstract}
We derive a general closed expression for the local pressure exerted onto the corrugated walls of a 
channel confining a fluid medium.
When the fluid medium is at equilibrium the local pressure is a functional of the shape of the walls.
It is shown that, due to the intrinsic non-local character of the interactions among the particles forming
the fluid, the applicability of approximate schemes such as the concept of a surface of tension or 
morphometric thermodynamics is limited to wall curvatures small compared to the range of particle-particle
interactions. 
\end{abstract}

\maketitle


\section{Introduction}

In a variety of scenarios, such as biological systems~\cite{Albers}, micro- and nano-fluidic 
circuitry~\cite{Duncombe2015}, and energy harvesting devices~\cite{Wang2012}, physical systems are 
composed of a fluid medium confined by means of solid or liquid interfaces. 
Clearly, the confining walls affect both the dynamics and the steady states of the fluid medium due to the 
additional interactions between the fluid phase and the boundaries. 
For example, droplet formation~\cite{Bakker1908}, electric charge accumulation in confined 
electrolytes~\cite{Frenkel1901}, and active particle accumulation in microchannels~\cite{Rothschild1963}
are driven by the effective interactions between the otherwise unbounded fluid medium and the confining
walls.
In order to grasp the effect of the presence of the confining walls on the dynamics of confined fluids, up
to now, the attention has been focused on the case of fluid media embedded between parallel plates or in 
channels with constant cross-sections. 
However, many experimentally relevant cases are not covered by these geometries. 
For example, porous materials, microfluidic devices, or biological scenarios, are characterized by larger
cavities that alternate with narrow bottlenecks.
This inhomogeneity in the confining space couples to the dynamics of the confined systems possibly leading to
novel scenarios. Indeed, electrolytes confined in varying section pores have been observed to undergo novel
regimes \cite{Siwy2008,Malgaretti2014,Keyser2015,Gracheva2017,Vinogradova2017} absent in the corresponding
unbound scenarios.
Similarly both passive \cite{Reguera2001,Bezrukov,Dagdug2012,Reguera2012,Marconi2015,Malgaretti2016} and 
active \cite{Di_Leonardo2010,Malgaretti2012,Ghosh2013,Dagdug2014,Malgaretti2017} colloidal particles as well
as polymers \cite{Muthukumar2008,Fazli2011,Sommer2013,Bianco2016,Datar2017} display confinement-induced 
dynamical regimes when embedded within varying section channels.  
In order to unravel the the general mechanisms responsible of these diverse phenomena, it is of primary
importance to understand the physical origin of the effective coupling between the fluid medium and the
confining walls. 
Accordingly, an insight into such a coupling can be provided by the knowledge of the dependence of local 
intensive thermodynamic quantities, e.g., pressure, on the local geometry of the confining walls. 
Indeed local imbalances in such quantities will lead to net local fluxes that are tuned by the geometry of
the confining walls. 
Alternatively, imposing the equilibrium condition, i.e., constancy of intensive thermodynamic
variables, will provide insight into the rearrangement of the fluid medium under inhomogeneous confining
conditions. 

In the following, we derive a framework capable of capturing the local dependence of thermodynamic variables
of the embedded fluid medium, and we derive a general expression for the local pressure exerted onto the
confinement. 
To that end, the fluid medium is described in terms of density functional theory (DFT), which is in principle
exact \cite{Evans1979} and for which numerous considerably precise approximation schemes have been developed
in the past \cite{Tarazona1984, Tarazona1985, Rosenfeld1989, Roth2002, Wittmann2016}.
Accordingly, our closed expression for the local pressure is exact within this framework.

\begin{figure}[t!]
   \includegraphics[width=8cm]{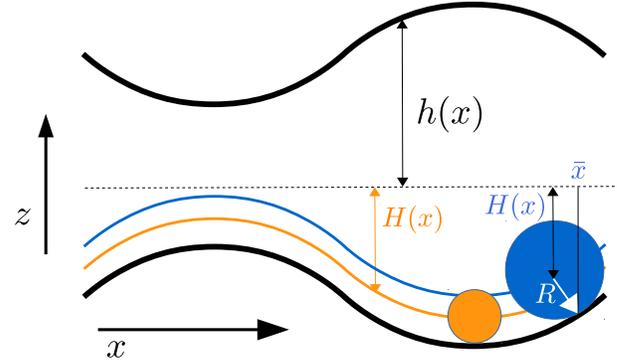}
   \caption{Cartoon of the system under study.
            Particles (depicted as discs) are confined to the interior of a channel whose upper and lower
            walls are located respectively at $z=h_+(x)$ and $z=-h_-(x)$. 
            Due to their radius $R_\alpha$, the volume accessible to particles of species $\alpha$ is given
            by $z\in[-H_\alpha^-(x),H_\alpha^+(x)]$.
            A particle of species $\alpha$ located at position $(x,\pm H_\alpha^\pm(x))$ touches the wall
            at point $(\bar{x}^\pm_\alpha(x,[h_\pm]), h_\pm(\bar{x}^\pm_\alpha(x,[h_\pm])))$.} 
   \label{fig:scheme}
\end{figure}

The structure of the text is the following. In Sec.~\ref{sec:gen-form} we define the model 
(Sec.~\ref{subsec:model}), set up the DFT approach (Sec.~\ref{subsec:DFT}), and derive a closed expression for
the local pressure (Sec.~\ref{subsec:workpressure}). 
In order to demonstrate the application of the general form of the local pressure derived in 
Sec.~\ref{sec:gen-form}, we discuss our results in the context of the concept of a surface of tension
and in comparison with the approach of morphometric thermodynamics in Sec.~\ref{sec:discussion}. 
Finally in Sec.~\ref{sec:conclusions} we give some concluding remarks.


\section{General formalism\label{sec:gen-form}}

\subsection{Model\label{subsec:model}}

As sketched in Fig.~\ref{fig:scheme}, consider in the three-dimensional Euclidean space an 
$s$-component mixture which is confined by two hard walls 
\begin{subequations}
\begin{align}
\mathcal{W}_+[h_+]:=&\{(x,y,z)\in\mathbb{R}^3\ |\ z>\phantom{-}h_+(x)\}\\
\mathcal{W}_-[h_-]:=&\{(x,y,z)\in\mathbb{R}^3\ |\ z<-h_-(x)\}\,, 
\end{align}
\end{subequations}
where the \textit{positive} real functions $h_\pm:\mathbb{R}\to(0,\infty)$ describe the wall shapes.
The particles of species $\alpha\in\{1,\dots,s\}$ possess a spherical hard core of radius $R_\alpha\geq0$, 
and, due to the hard walls $\mathcal{W}_+[h_+]$ and $\mathcal{W}_-[h_-]$, they are confined to the interior
of the accessible volume $\mathcal{V}_\alpha[h]:=\{(x,y,z)\in\mathbb{R}^3\ |\ -H^-_\alpha(x,[h_-])
<z<H^+_\alpha(x,[h_+])\}$ with the positive real functions $H^\pm_\alpha[h_\pm]:\mathbb{R}\to(0,\infty)$.
Here and in the following $h:=(h_+,h_-)$ is an abbreviation of the set of both wall shape functions $h_+$
and $h_-$.
In order to have the functions $H^\pm_\alpha[h_\pm]$, which describe the boundary 
$\partial\mathcal{V}_\alpha[h]$ of the accessible volume $\mathcal{V}_\alpha[h]$, 
being well-defined, the radii of curvature $R^\pm_\text{w}(x,[h_\pm])
:=(1+h_\pm'(x)^2)^{3/2}/h_\pm''(x)$ of the walls are required to be larger than the radius of all particles:
$|R^\pm_\text{w}(x)|>R_\alpha$ for all $x\in\mathbb{R}$ and $\alpha\in\{1,\dots,s\}$.
Besides the hard-core interactions, additional fluid-fluid and fluid-wall interactions can be present.
The structure of the fluid is described in terms of the number density profiles $\rho_\alpha$ of species 
$\alpha\in\{1,\dots,s\}$, which, due to the translational invariance of the system along the $y$-direction,
can be assumed to depend only on $x$ and $z$. 
The hard-core interaction precludes particles of species $\alpha\in\{1,\dots,s\}$ to be found outside the
accessible volume $\mathcal{V}_\alpha$, i.e., $\rho_\alpha(x,z)=0$ for $(x,y,z)\not\in 
\mathcal{V}_\alpha[h]$.


\subsection{Density functional theory\label{subsec:DFT}}

Without restriction of generality the grand-canonical density functional describing the fluid is given by
\begin{align}
   \beta\Omega\left[\set{\rho},h\right] 
   =
   \mathcal{A}\left[\set{\rho},\set{H}^+[h_+],\set{H}^-[h_-],h\right]
   \label{eq:GC0}
\end{align}
with the auxiliary functional
\begin{widetext}
\begin{align}
   \mathcal{A}\left[\set{\rho},\set{J}^+,\set{J}^-,h\right]
   =
   L_y \int\!\d x \sum_{\alpha=1}^s \int\limits_{-J^-_\alpha(x)}^{J^+_\alpha(x)}\!\!\!\!\!\d z\; 
   \rho_\alpha(x,z)
   \left[f_\alpha\left(\rho_\alpha(x,z)\right) - \beta\mu_\alpha + \beta V_\alpha(x,z,[h])\right]
   + \beta F^\text{ex}\left[\set{\rho},\set{J}^+,\set{J}^-\right],
   \label{eq:A}
\end{align}
\end{widetext}
where $\set{\rho}=(\rho_1,\dots,\rho_s)$ denotes the set of all density profiles and similarly for other
quantities, and $L_y$ is the thickness of the system along the $y$ direction.
In Eq.~(\ref{eq:A}), $V_\alpha(x,z,[h])$ and $\mu_\alpha$ are respectively the particle-wall interaction in
excess to the hard-core interaction for particles of species $\alpha\in\{1,\dots,s\}$ and the chemical
potential.
Note that $V_\alpha(x,z,[h])$ depends explicitly on the shapes $h_+$ and $h_-$ of the confining
walls.
The reference system is described by the free energy density per particle $f_\alpha(\rho_\alpha)$, which
equals exactly $\ln(\rho_\alpha\Lambda^3_\alpha)-1$ with the thermal wave length $\Lambda_\alpha$ for
the case of an ideal gas, but which also allows for a description within a local density approximation (LDA).
In the present work the free energy functional $F^\text{ex}$ in excess to the reference functional is chosen
of the rather general form
\begin{widetext}
\begin{align}
   &\beta F^\text{ex}\left[\set{\rho},\set{J}^+,\set{J}^-\right]
   \notag\\
   = 
   &-L_{y}\!\sum_{n=2}^\infty\frac{1}{n!} \sum_{\alpha_1,\dots,\alpha_n=1}^s 
   \int\!\d x_1\!\!\!\!\!\!\!\!\int\limits_{-J^-_{\alpha_1}(x_1)}^{J^+_{\alpha_1}(x_1)}\!\!\!\!\!\!\!\d z_1
   \dots
   \int\!\d x_n\!\!\!\!\!\!\!\!\int\limits_{-J^-_{\alpha_n}(x_n)}^{J^+_{\alpha_n}(x_n)}\!\!\!\!\!\!\!\d z_n\;
   c_{\alpha_1\dots\alpha_n}^{(0)}(x_1,z_1,\dots,x_n,z_n)
   \rho_{\alpha_1}(x_1,z_1)\cdots\rho_{\alpha_n}(x_n,z_n).
   \label{eq:Fex}
\end{align}
\end{widetext}
All weighted density approximations (WDA), including the fundamental measure theories (FMT), are of this 
form.
However, gradient expansions, such as the well-known square-gradient approximation within the Cahn-Hilliard
approach, cannot be represented in the form of Eq.~(\ref{eq:Fex}).

Given some wall shapes $h$, the equilibrium number density profiles $\rho^\text{eq}[h]$ are 
solutions of the Euler-Lagrange equations
\begin{align}
   0 
   &=
   \frac{\delta\beta\Omega}{\delta\rho_\alpha(x,z)}[\set{\rho}^\text{eq}[h],h] 
   \notag\\
   &= 
   \frac{\delta\mathcal{A}}{\delta\rho_\alpha(x,z)}[\set{\rho}^\text{eq}[h],\set{H}^+[h_+],\set{H}^-[h_-],h] 
   \label{eq:ELe}
\end{align}
for all $\alpha\in\{1,\dots,s\}$ and $(x,y,z)\in\mathcal{V}_\alpha[h]$.


\subsection{Deformation work and local pressure\label{subsec:workpressure}}

When the fluid medium is at equilibrium, or in some non-equilibrium steady state, 
the conditions of the system, composed of the walls and the fluid medium, are such that the 
state of the fluid medium, represented by the density profiles $\set{\rho}$, is determined by the shapes
of the walls $h_\pm$ alone: $\set{\rho}=\set{\rho}^*[h]$.
In such a scenario one obtains from Eq.~(\ref{eq:GC0}) the effective wall Hamiltonian on the function space
of all wall shapes $h=(h_+, h_-)$
\begin{align}
   \beta\Xi[h] := \beta\Omega\left[\set{\rho}^*[h],h\right]
   \label{eq:Xigeneral}
\end{align}
that represents the energy of the system for given wall shapes $h$.

In the following, we restrict ourselves to the case of the fluid medium being at equilibrium, i.e., 
$\set{\rho}^*=\set{\rho}^\text{eq}$, where the equilibrium state is given by Eq.~(\ref{eq:ELe}). 
In such a condition Eq.~(\ref{eq:Xigeneral}) reads 
\begin{align}
   \beta\Xi[h] = \beta\Omega\left[\set{\rho}^\text{eq}[h],h\right],
   \label{eq:Xi}
\end{align}
and the degrees of freedom of the fluid medium are accounted for in terms of the equilibrium state 
$\set{\rho}^\text{eq}[h]$. 

Accordingly, as is shown in detail in Appendix~\ref{app:deltaXi}, the work needed to deform the walls 
$h=(h_+,h_-)$ by $\delta h=(\delta h_+,\delta h_-)$ is given by
\begin{widetext}
\begin{align}
   \delta\beta\Xi\left[h,\delta h\right] 
   &= L_y\int\!\d x' \sum_{\alpha=1}^s\sum_{t\in\{\pm\}}\!\!\Bigg[
   -\partial_{x'}K^t_\alpha(x',[h_t])\left.\rho_\alpha^\text{eq}(x,tH^t_\alpha(x,[h_t]),[h])^2
   f_\alpha'(\rho_\alpha^\text{eq}(x,tH^t_\alpha(x,[h_t]),[h]))\right|_{x=K^t_\alpha(x',[h_t])}
   \notag\\
   &\phantom{= L_y\int\!\d x' \sum_{\alpha=1}^s\sum_{t\in\{\pm\}}\!\!\Bigg[}
   + \int\!\d x\int\limits_{-H^-_\alpha(x,[h_-])}^{H^+_\alpha(x,[h_+])}\!\!\!\d z\;
     \rho_\alpha^\text{eq}(x,z,[h])\frac{\delta\beta V_\alpha}{\delta h_t(x')}(x,z,[h])\Bigg]
   \ \delta h_t(x'),
   \label{eq:tot-var3}
\end{align}
where (compare Eq.~(\ref{eq:def-K}) in Appendix~\ref{app:deltaXi})
\begin{align}
   K^\pm_\alpha(x',[h_\pm]) := x' + \frac{R_\alpha h_\pm'(x')}{\sqrt{1+h_\pm'(x')^2}}.
   \label{eq:def-K0}
\end{align}

Equation~(\ref{eq:tot-var3}) is an expression of the work $\beta\Xi[h,\delta h]$ required to 
deform the walls, whose shape is described by $h=(h_+,h_-)$, such that $h$ is changed by 
$\delta h$.
It is of the form
\begin{align}
   \delta\Xi\left[h,\delta h\right]
   = -L_y\int\!\d x\left[\beta P^+_z(x,[h])\ \delta h_+(x) + \beta P^-_z(x,[h])\ \delta h_-(x)\right]
   \label{eq:tot-var4}
\end{align}
where
\begin{align}
   \beta P^t_z(x,[h]) 
   &=
   \sum_{\alpha=1}^s\Bigg[
   \partial_{x}K^t_\alpha(x,[h_t])\left.\rho_\alpha^\text{eq}(x',tH^t_\alpha(x',[h_t]),[h])^2
   f_\alpha'(\rho_\alpha^\text{eq}(x',tH^t_\alpha(x',[h_t]),[h]))\right|_{x'=K^t_\alpha(x,[h_t])}\
   \notag\\
   &\phantom{= \sum_{\alpha=1}^s\Bigg[}
   - \int\!\d x'\int\limits_{-H^-_\alpha(x',[h_-])}^{H^+_\alpha(x',[h_+])}\!\!\!\d z'\;
     \rho_\alpha^\text{eq}(x',z',[h])\frac{\delta\beta V_\alpha}{\delta h_t(x)}(x',z',[h])\Bigg]
   \label{eq:P1}
\end{align}
is the local force along the positive ($t=+$) or negative ($t=-$) $z$-direction per unit area projected onto the axis 
of the channel the fluid exerts on the walls at $(x,y,th_t(x))$.
Using as reference system in Eq.~(\ref{eq:A}) the ideal gas with free energy per particle 
$f_\alpha(\rho_\alpha) = \ln(\rho_\alpha\Lambda_\alpha^3)-1$ one obtains $\rho_\alpha^2f'_\alpha(\rho_\alpha)=
\rho_\alpha$, so that Eq.~(\ref{eq:P1}) simplifies to
\begin{align}
   \beta P^t_z(x,[h])
   &= 
   \sum_{\alpha=1}^s\Bigg[
   \partial_xK^t_\alpha(x,[h_t])\rho_\alpha^\text{eq}(K^t_\alpha(x,[h_t]),tH^t_\alpha(K^t_\alpha(x,[h_t]),
   [h_t]),[h])
   \notag\\
   &\phantom{= \sum_{\alpha=1}^s\Bigg[} 
   - \int\!\d x'\int\limits_{-H^-_\alpha(x',[h_-])}^{H^+_\alpha(x',[h_+])}\!\!\!\d z'\;
     \rho_\alpha^\text{eq}(x',z',[h])\frac{\delta\beta V_\alpha}{\delta h_t(x)}(x',z',[h])\Bigg].
   \label{eq:gen-P}
\end{align}
\end{widetext}

The rotation invariance of $P^t_z$ (see Appendix \ref{app:Pz}) allows one to determine the local force 
$P^t_x(x,[h])$ along the $x$-direction per unit area projected onto the axis of the channel the fluid exerts
on the walls at $(x,y,th_t(x))$.
By definition the force exerted on a surface element of the wall at point $(x,y,th_t(x))$ with rectangular
projection of size $\d x$ and $L_y$ onto the $x$-$y$ plane is given by $(P^t_x(x,[h]), 
tP^t_z(x,[h]))L_y\d x$.
As the force is a vectorial quantity, it transforms upon rotation $T(\varphi)$ with the rotation matrix
$\mathbb{T}(\varphi)$:
\begin{align}
   \left(\begin{array}{c}
      P^t_x(\tilde{x},[\tilde{h}]) \\
      tP^t_z(\tilde{x},[\tilde{h}])
   \end{array}\right)
   L_y\d\tilde{x}
   =
   \mathbb{T}(\varphi)\left(\begin{array}{c}
      P^t_x(x,[h]) \\
      tP^t_z(x,[h])
   \end{array}\right)
   L_y\d x.
   \label{eq:force}
\end{align}
Substituting $z=th_t(x)$ in Eq.~(\ref{eq:rotation}) and forming the differential of the first component one 
obtains the expression $\d\tilde{x}=(\cos\varphi - \sin\varphi th_t'(x))\d x$.
Using this relation as well as Eq.~(\ref{eq:rotinv}) one infers from the second component in 
Eq.~(\ref{eq:force}) the expression $P^t_x(x,[h]) = -h_t'(x) P^t_z(x,[h])$, and hence the force onto
a surface element of the wall (see Eq.~(\ref{eq:force})) is given by
\begin{align}
   \left(\begin{array}{c}
      P^t_x(x,[h]) \\
      tP^t_z(x,[h])
   \end{array}\right)
   L_y\d x
   &= 
   P^t_z(x,[h])
   \left(\begin{array}{c}
      -h_t'(x) \\
      t
   \end{array}\right)
   L_y\d x
   \notag\\
   &= 
   P^t_z(x,[h])
   \vec{n}_t(x,[h_t]) \d\mathcal{A},
\end{align}
where $\vec{n}_t(x,[h_t])=(-h_t'(x),t)/\sqrt{1+h_t'(x)^2}$ is the outer normal vector of the wall at 
$(x,y,h_t(x))$ and $\d\mathcal{A}=L_y\sqrt{1+h_t'(x)^2}\d x$ is the size of the surface element.
Therefore, in agreement with the intuitive expectation that a fluid in equilibrium cannot sustain shear, the
force the fluid medium is exerting onto the walls acts in normal direction.
Moreover, the quantities $P^t_z$ in Eqs.~(\ref{eq:P1}) and (\ref{eq:gen-P}) represent the force per wall area,
i.e., the \textit{local pressure}.
Therefore Eqs.~(\ref{eq:P1}) and (\ref{eq:gen-P}) can be seen as a local version of the so-called
``contact theorems'' which, so far, consider effectively averages of the wall density, either
due to the symmetry at regularly formed walls \cite{Lebowitz1960, Henderson1979, Henderson1983, Podgornik1993,
Trizac2015} or as explicit averages along arbitrarily-shaped walls \cite{Upton1998}.


\section{Discussion\label{sec:discussion}}

\subsection{Sum rules}

Some general statements concerning the local pressures $P_z^\pm(x,[h])$ can inferred directly from the 
expression for the mechanical work, Eq.~(\ref{eq:tot-var4}), without the need to determine the equilibrium
density profiles $\varrho_\alpha^\text{eq}$. 
The work required to induce the deformation $\delta h=(\delta h_+,\delta h_-)$ of the channel walls in
general reads
\begin{equation}
   \delta\Xi = 
   -p\delta\mathcal{V} + \bar{\gamma}_+\delta\mathcal{A}_+ + \bar{\gamma}_-\delta\mathcal{A}_-,
\label{eq:tot-work}
\end{equation}
where $\delta\mathcal{V}$ and $\delta\mathcal{A}_\pm$ are respectively the changes in volume and in surface
areas due to the deformation $\delta h$ of the walls, $p$ is the bulk pressure, and $\bar{\gamma}_\pm$ are
the associated interfacial tensions. 
For the special case of deformations $\delta h$ leaving the size of the surface areas unchanged,
\begin{equation}
\begin{cases}
 \delta\mathcal{V}&\neq 0\\
 \delta \mathcal{A}_\pm&=0,
\end{cases}
\end{equation}
the total work depends solely on the ``bulk'' contribution $\sim\delta\mathcal{V}$ on the right-hand side of
Eq.~(\ref{eq:tot-work}). 
Comparing Eq.~(\ref{eq:tot-work}) to Eq.~(\ref{eq:tot-var4}) under these conditions one infers the sum rule
\begin{align}
   \int\!\d x (P^\pm_z(x,[h])-p)\ \delta h_\pm(x)=0\,.
   \label{eq:sumrule1}
\end{align}
In particular, when the walls are deformed uniformly, i.e., $\delta h_\pm(x)$ does not depend on the position 
$x$ along the channel, one obtains from Eq.~(\ref{eq:sumrule1}) the sum rule
\begin{align}
   \int\!\d x (P^\pm_z(x,[h]) - p) = 0
   \label{eq:sumrule2}
\end{align}
i.e., the \textit{average} pressure equals the bulk pressure, $p$.

Alternatively, when the deformation $\delta h$ is such that the surface areas are modified with the volume 
left unchanged,
\begin{equation}
\begin{cases}
 \delta\mathcal{V}&=0\\
 \delta \mathcal{A}_\pm&\neq0,
\end{cases}
\end{equation}
the mechanical work is due to the terms $\sim\delta\mathcal{A}_\pm$ on the right-hand side of 
Eq.~(\ref{eq:tot-work}). 
Accordingly, comparing Eq.~(\ref{eq:tot-work}) to Eq.~(\ref{eq:tot-var4}) one infers the sum rule
\begin{equation}
   \int\!\d x\!\!\left[\!P^\pm_z(x,[h]) \!-\! 
   \bar{\gamma}_\pm\frac{\partial}{\partial x}\!\!\left(\dfrac{h'_\pm(x)}{\sqrt{1+h'_\pm(x)^2}}\right)
   \!\right]\!\delta h_\pm(x)\!=\!0.
   \label{eq:sumrule3}
\end{equation}
In the case of a planar channel, $h(x)=(h_0,h_0), h_0>0$, the second term in the integrand of 
Eq.~(\ref{eq:sumrule3}) vanishes.
Hence the mechanical work needed to deformation a flat channel is at least quadratic in the deformations
$\delta h$, which is not captured by the linear response functions $P^\pm_z(x,[h])$.


\subsection{Non-interacting particles}

An application of the main result Eqs.~(\ref{eq:P1}) and (\ref{eq:gen-P}) to calculate the local pressure
requires knowledge of the the equilibrium density profiles $\rho^\text{eq}_\alpha$, usually from solving
the Euler-Langrange equations~(\ref{eq:ELe}).
In actual applications the latter task can be a challenging, typically numerical, problem.
However, in order to highlight some important consequences of Eq.~(\ref{eq:gen-P}) the following discussion
is focusing on computationally simple cases of a single particle species ($s=1$) in a semi-infinite system
($h_-(x)=\infty$) for a vanishing excess external potential ($V_\alpha=0$).
Under these conditions only the local pressure onto the upper wall is relevant so that Eq.~(\ref{eq:gen-P}) 
takes the form
\begin{align}
   \beta P_z(x,[h]) =
   &\ 
   \left(1 + \frac{R}{R_\text{w}(x,[h])}\right)
   \label{eq:gen-P1}\\
   &\ 
   \times\rho^\text{eq}(K(x,[h]),H(K(x,[h]),[h]),[h]),
   \notag
\end{align}
where Eq.~(\ref{eq:k_alpha}) has been used and the indices $\alpha$ and $+$ have been suppressed.

As a first case consider a single-component fluid of non-interacting particles ($U=0$), for which the
equilibrium number density $\rho^\text{eq}(x,z) = \beta p$, $p$ being the pressure of the particle
reservoir, is uniform inside the accessible volume. 
Upon rearrangement one obtains from Eq.~(\ref{eq:gen-P1})
\begin{align}
   \beta p - \beta P_z(x,[h]) = \beta pR\frac{1}{-R_\text{w}(x,[h])}.
   \label{eq:Laplace1_nonint}
\end{align}
The fraction on the right-hand side of Eq.~(\ref{eq:Laplace1_nonint}) represents the wall curvature which is
positive (negative) at concave (convex) portions of the wall when viewed from inside the fluid.
Equation~(\ref{eq:Laplace1_nonint}) shows that point-like particles ($R=0$) imply $p=P_z(x,[h])$,
i.e., the local pressure, $P_z(x,[h])$, always equals the bulk pressure, $p$. 
In contrast, for finite-sized particles ($R>0$) such as hard spheres the local pressure is not constant at
walls of non-uniform curvature (see Eq.~(\ref{eq:Laplace1_nonint})), and it exceeds the bulk pressure
when $R_\text{w}>0$ (convex wall) whereas the opposite holds for  $R_\text{w}<0$ (concave wall).
Next we compare our result Eq.~(\ref{eq:Laplace1_nonint}) to the concept of \textit{surface of 
tension} and to \textit{morphometric thermodynamics}. 

\paragraph*{Surface of tension.}
At first glance, Eq.~(\ref{eq:Laplace1_nonint}) is similar to Laplace's law. 
However to make the comparison more strict one has to locate the Surface Of Tension (SOT) and to specify
the corresponding interfacial tension in Eq.~(\ref{eq:Laplace1_nonint}). 
To this end, we note that, in general, the interfacial tension is defined as
\begin{align}
   \beta\gamma= 
   \frac{\beta \Omega[\rho^\text{eq}[h],[h]] + \beta p \tilde{V}}{\tilde{A}}
   \label{eq:def-gamma}
\end{align}
where $\Omega[\rho^\text{eq}[h],[h]]$ is the grand-canonical potential of the system (enclosing an 
accessible volume $V$), $-p \tilde{V}$ is the grand-canonical potential of a uniform system
with bulk pressure $p$ inside the conventional volume $\tilde{V}$, and $\tilde{A}$ is the area of the 
dividing surface.
For a cylindrical wall with uniform radius of curvature $|R_\text{w}|$, the interfacial tension with respect
to the dividing surface of radius $\widetilde{R}>0$ is 
\begin{align}
   \beta\gamma(\tilde{R}) = 
   \frac{-\beta p(|R_\text{w}|-R)^2 + \beta p\widetilde{R}^2}{2\widetilde{R}}
   \label{eq:gammaRtilde}
\end{align}
and therefore, by subtracting Eq.~(\ref{eq:gammaRtilde}) from Eq.~(\ref{eq:Laplace1_nonint}) one obtains:
\begin{align}
   \beta p - \beta P_z - \frac{\beta\gamma(\widetilde{R})}{\widetilde{R}}
   =
   \frac{\beta pR}{|R_\text{w}|} + 
   \frac{\beta p}{2}\left(\!\!\left(\frac{|R_\text{w}|-R}{\widetilde{R}}\right)^2\!\! - 1\right).
   \label{eq:sot}
\end{align}
The SOT is defined by that value $\widetilde{R}=R_\text{SOT}$ for which the right-hand
side of Eq.~(\ref{eq:sot}) vanishes, i.e., with respect to which Laplace's law is valid exactly 
\cite{RowlinsonWidom}.
The SOT is a function of the wall curvature $1/|R_\text{w}|$ and of the size of the fluid particles $R$:
\begin{align}
   R_\text{SOT} = \frac{|R_\text{w}|-R}{\dps\sqrt{1 - \frac{2R}{|R_\text{w}|}}}.
   \label{eq:Rsot}
\end{align}
Notice that the SOT is defined only for $R<2|R_\text{w}|$, whereas Eq.~(\ref{eq:gen-P1}) has been derived
under the weaker constraint $R<|R_\text{w}|$.
Using Eq.~(\ref{eq:Rsot}) in Eq.~(\ref{eq:gammaRtilde}) one obtains the curvature dependence of the
interfacial tension $\beta\gamma(R_\text{SOT})$ at the SOT.

\paragraph*{Morphometric thermodynamics.}
In the past years several publications on the properties of fluids in contact with geometrically shaped
walls advocated for an approach termed ``Morphometric Thermodynamics'' (MT) \cite{Koenig2004,Roth2006,
HansenGoos2007,Evans2014}.
The basic hypothesis of MT is that any thermodynamic quantity can be expressed as a linear combination
of four geometric measures namely: volume, surface area, integrated mean curvature, and integrated Gaussian
curvature, with \emph{intensive} coefficients, which depend only on the thermodynamic state of
the system \cite{Koenig2004}.
Whereas the framework of the surface of tension allows for a curvature dependence of the interfacial tension,
the intensive coefficients within MT are strictly independent of the geometrical features of the system.
If applicable, MT is a very efficient method to characterize fluids in complicated confinements.
However, in recent years several indications have been found that MT is in general merely an 
approximation \cite{Laird2012,HansenGoos2014,Reindl2015}.
For the present fluid of non-interacting particles the interfacial tension $\gamma_\text{geo}$ of a planar
wall with respect to the geometrical wall surface is given by $\beta\gamma_\text{geo}=\beta pR$, as the fluid
particles cannot approach the wall closer than $R$.
Hence Eq.~(\ref{eq:Laplace1_nonint}) takes the morphometric form
\begin{align}
   \beta p - \beta P_z(x,[h]) = \frac{\beta\gamma_\text{geo}}{-R_\text{w}(x,[h])}.
   \label{eq:Laplace_nonint_MT}
\end{align}
Note, however, that here the MT result is achieved for planar interfacial tensions only if the dividing
surface is the geometrical wall surface. 
Any other choice of the dividing surface will lead to additional terms in 
Eq.~(\ref{eq:Laplace_nonint_MT}) that are not in agreement with the MT approach.


\subsection{Interacting particles}

Whereas non-vanishing particle-particle interactions ($U\not=0$) are typically too complicated to be treated
analytically, the effect of tuning on the interaction is accessible in the limit of small interactions.
Indeed, introducing the auxiliary interaction potential $U_\eta(\vec{r}) := \eta U(\vec{r})$ for 
$\eta\in[0,1]$ a closed expression for the local pressure can be obtained in the limit $\eta\to0$ via a 
Taylor-expansion of Eq.~(\ref{eq:gen-P1}) about the non-interacting case, $\eta=0$.
At first order in $\eta$, from Eq.~(\ref{eq:gen-P1}), one obtains
\begin{align}
   \Big(\frac{\d}{\d\eta}&\beta P_z(x,[U_\eta,h])\Big)\Big|_{\eta=0} =
   \left(1 + \frac{R}{R_\text{w}(x,[h])}\right)
   \label{eq:d_deta}\\
   &\times\!\!\Big(\frac{\d}{\d\eta}\rho^\text{eq}(K(x,[h]),H(K(x,[h]),[h]),[U_\eta,h])\Big)\Big|_{\eta=0}
   \!\!.
   \notag
\end{align}
Upon taking the derivative of the Euler-Lagrange equation~(\ref{eq:ELe}) with respect to the interaction
potential and using the general relation \cite{Evans1979}
\begin{align}
   \frac{\delta F^\text{ex}[\rho,U]}{\delta U(\vec{r},\vec{r'})} =
   \frac{1}{2}\rho(\vec{r})\rho(\vec{r'})g(\vec{r},\vec{r'},[\rho,U])
\end{align}
one can derive 
\begin{align}
   &\Big(\frac{\d}{\d\eta}\rho^\text{eq}(K(x,[h]),H(K(x,[h]),[h]),[U_\eta,h])\Big)\Big|_{\eta=0}
   \notag\\
   =&-(\beta p)^2\int\d x'\int\d y'\int\limits_{-\infty}^{H(x',[h])}\d z'\beta U(\vec{r}-\vec{r'}),
   \label{eq:drhodeta}
\end{align}
where $\vec{r}=(K(x,[h]),y,H(K(x,[h]),[h]))$ and $\vec{r'}=(x',y',z')$.
In the light of Eq.~(\ref{eq:drhodeta}) it is plausible that, due to the integral over the accessible volume,
one obtains a complicated non-local dependence of the local pressure on the shape of the walls upon 
switching on a non-local interaction between the fluid particles.
Moreover, due to the the non-local dependence of the right-hand side of Eq.~(\ref{eq:d_deta}) on the
shape of the contact surface (see Eq.~(\ref{eq:drhodeta})), for walls with non-constant curvature 
$1/R_\text{w}(x,[h])$ and for non-local interactions $U(\vec{r})$ the pressure difference $p[U] - 
P_z^\pm[x,U,h]$ is a functional of the whole wall shape $h$ and not only a function of the local wall
curvature $1/R_\text{w}^\pm(x,[h_\pm])$.
Hence, for arbitrary wall shapes and non-local interactions, neither MT nor the concept of a SOT are applicable
exactly.
The range of wall curvatures, for which MT and the SOT can be applied as approximations, is decreasing upon
increasing the range of the interactions inside the fluid.

\paragraph*{Morphometric thermodynamics.}
For the case of uniform curvature $1/|R_\text{w}|$ considered above and for the switching on
of a square-well or square-shoulder interaction $U(\vec{r})=U_0\Theta(R_0-|\vec{r}|)$ with strength $U_0$
and range $R_0\neq0$ one obtains
\begin{align}
   &\Big(\frac{\d}{\d\eta}\Big(\beta p[U_\eta] - \beta P_z[U_\eta] - 
   \frac{\beta\gamma_\text{geo}[U_\eta]}{|R_\text{w}|}\Big)\Big)\Big|_{\eta=0}
   \notag\\
   =&\frac{\pi(\beta p)^2\beta U_0 R_0^6}{384|R_\text{w}|^3} + 
   R_0^3\mathcal{O}^5\Big(\frac{R_0}{|R_\text{w}|},\frac{R}{|R_\text{w}|}\Big).
   \label{eq:d_detaLaplace}
\end{align}
Whereas Eq.~(\ref{eq:d_detaLaplace}) can be rewritten in the form of Laplace's law by means of an 
appropriate definition of the SOT, it is impossible to achieve the morphometric form for $\beta U_0\not=0$
and $R_0\neq0$.
The latter statement is justified as follows: the non-vanishing term $\sim 1/|R_\text{w}|^3$ on the
right-hand side is forbidden within MT approach since it is of higher power in the wall curvature.
Absorbing this term into an interfacial tension with respect to some dividing surface (characterized by 
$\widetilde{R}$) different from the geometrical wall surface is possible, however, this would generate an 
additional term $\sim 1/|\widetilde{R}|^2$, which is also forbidden within MT.
One can conclude that for switching on a non-local interaction, i.e., with $R_0\neq0$, MT cannot be expected
to be strictly applicable and the best approximation of MT is achieved with the geometrical wall surface as
dividing interface, within which corrections to MT are $\sim 1/|R_\text{w}|^3$ and $\sim 1/|\widetilde{R}|^2$
otherwise.


\section{Conclusions\label{sec:conclusions}}

We have derived a general formula for the dependence of the
local pressure onto the walls of a channel with varying cross-section (see Eq.~(\ref{eq:P1})). 
Such a closed formula has been extracted from the mechanical work (Eq.~(\ref{eq:Xi})) needed to induce a local 
deformation of the shape of the wall confining a fluid medium and under the condition of equilibrium
of the fluid medium. In such a scenario we have identified the contribution to the mechanical work due to the
local pressure and we have derived a general formula, Eq.~(\ref{eq:P1}), for the dependence of the local
pressure onto the walls of a channel with varying cross-section. 
This result applies to the wide class of fluid media which can be described by density functionals with excess
contributions of the form Eq.~(\ref{eq:Fex}).

In the case of purely hard walls, the local pressure depends only on the curvatures of the walls and on
the densities at the surfaces of contact.
As expected, for point-like non-interacting particles the mechanical work is due solely to the bulk
pressure. In contrast, for finite-sized particle, such as hard-spheres, and/or interacting particles, we have
identified the contributions to the mechanical work stemming from the interfacial tension. 
In particular, for non-interacting finite-sized particles confined by a constant-curvature wall (e.g., a
semi-cylinder) we have derived a closed formula for the local pressure. 
This has allowed us to derive the dependence of the interfacial tension upon the choice of the dividing 
surface (encoded in $\widetilde{R}$ in Eq.~(\ref{eq:gammaRtilde})), and to identify the Surface Of 
Tension (SOT) as the dividing surface for which the Laplace law is recovered. 
Interestingly the SOT does not coincide with the wall surface. 
For non-interacting particles our result is in agreement with Morphometric Thermodynamics (MT)
\cite{Koenig2004}.

When the fluid particles are interacting among themselves the expression for the local pressure becomes more
involved. 
However, an insight into the corrections to MT can be derived for weakly interacting particles, for which
the expression for the local pressure can be obtained by expanding about the non-interacting case. 
In such a scenario, our results show that the MT approach breaks down (Eq.~(\ref{eq:d_detaLaplace})) in that
additional contributions with a non-MT form appear as corrections to the Laplace equation. 

Finally, for arbitrary wall shapes and non-local interactions the local pressure does not depend
locally on the wall curvature so that schemes such as the concept of a surface of tension or morphometric
thermodynamics are merely approximations, whose applicability is limited to ranges of small wall curvatures,
which decrease upon increasing the range of the interactions inside the fluid.


\section*{Acknowledgments}

PM thanks Erio Tosatti for pinpointing the problem under study and for useful discussions.
Moreover, we are grateful to S.\ Dietrich for valuable comments.


\appendix

\begin{widetext}

\section{Derivation of Equation~(\ref{eq:tot-var3})\label{app:deltaXi}}

According to Eq.~(\ref{eq:Xi}) in Sec.~\ref{subsec:workpressure} of the main text, the work needed to deform
the walls $h=(h_+,h_-)$ by $\delta h=(\delta h_+,\delta h_-)$ is given by
\begin{align}
   \delta\beta\Xi\left[h,\delta h\right] 
   &= \int\!\d x\;\Bigg( 
   \sum_{\alpha=1}^s\int\limits_{-J^-_\alpha(x)}^{J^+_\alpha(x)}\!\!\!\d z\;
      \frac{\delta\mathcal{A}}{\delta\rho_\alpha(x,z)}[\set{\rho}^\text{eq}[h],\set{J}^+,\set{J}^-,h]\
      \delta\rho_\alpha(x,z,[h,\delta h]) 
   \notag\\
   &\phantom{= \int\!\d x\;\Bigg(}
   + \sum_{\alpha=1}^s\sum_{t\in\{\pm\}}\!\!\Bigg(\frac{\delta\mathcal{A}}{\delta J^t_\alpha(x)}
     [\set{\rho}^\text{eq}[h],\set{J}^+,\set{J}^-,h] \delta J^t_\alpha(x)
   + \frac{\delta\mathcal{A}}{\delta h_t(x)}[\set{\rho}^\text{eq}[h],\set{J}^+,\set{J}^-,h]\ \delta h_t(x)
   \Bigg)\Bigg|_{\set{J}^\pm=\set{H}^\pm[h_\pm]}
   \notag\\
   &= \int\!\d x \sum_{\alpha=1}^s\sum_{t\in\{\pm\}}\!\!\Bigg[\Bigg(
   L_y \rho_\alpha^\text{eq}(x,tJ^t_\alpha(x),[h])\Big(f_\alpha(\rho_\alpha^\text{eq}(x,tJ^t_\alpha(x),[h])) 
   - \beta\mu_\alpha+\beta V_\alpha(x,tJ^t_\alpha(x),[h])\Big) 
   \notag\\
   &\phantom{= \int\!\d x \sum_{\alpha=1}^s\sum_{t\in\{\pm\}}\!\!\Bigg[\Bigg(}
   + \frac{\delta\beta F^\text{ex}}{\delta J^t_\alpha(x)}[\set{\rho}^\text{eq}[h],\set{J}^+,\set{J}^-]\Bigg)\ 
   \delta J^t_\alpha(x)
   \notag\\
   &\phantom{= \int\!\d x \sum_{\alpha=1}^s\sum_{t\in\{\pm\}}\!\!\Bigg[\Bigg(}
   + L_y \int\limits_{-J^-_\alpha(x)}^{J^+_\alpha(x)}\!\!\!\d z\;
      \rho_\alpha^\text{eq}(x,z,[h])\int\!\d x'\;\frac{\delta\beta V_\alpha}{\delta h_t(x')}(x,z,[h])\ 
      \delta h_t(x')\Bigg]
   \Bigg|_{\set{J}^\pm=\set{H}^\pm[h_\pm]},
   \label{eq:deltaXi}
\end{align}
where Eqs.~(\ref{eq:A}) and (\ref{eq:ELe}) have been used.
In particular, the Euler-Lagrange equation~(\ref{eq:ELe}) can be written in the form
\begin{equation}
   L_y \Big(f_\alpha(\rho_\alpha^\text{eq}(x,z,[h])) - \beta\mu_\alpha + \beta V_\alpha(x,z,[h])\Big)
   =-L_y \rho_\alpha^\text{eq}(x,z,[h]) f_\alpha'(\rho_\alpha^\text{eq}(x,z,[h]))
   -\frac{\delta\beta F^\text{ex}}{\delta \rho_\alpha(x,z)}[\set{\rho}^\text{eq}[h],\set{H}^+[h_+],
   \set{H}^-[h_-]]
\end{equation}
so that Eq.~(\ref{eq:deltaXi}) is given by
\begin{align}
   \delta\beta\Xi\left[h,\delta h\right] =
   &\ 
   \int\!\d x \sum_{\alpha=1}^s\sum_{t\in\{\pm\}}\!\!\Bigg[\delta H^t_\alpha(x,[h_t,\delta h_t])\Bigg(
   -L_y\rho_\alpha^\text{eq}(x,tH^t_\alpha(x,[h_t]),[h])^2f_\alpha'(\rho_\alpha^\text{eq}(x,
   tH^t_\alpha(x,[h_t]),[h]))
   \notag\\
   &\ 
   -\rho_\alpha^\text{eq}(x,tH^t_\alpha(x,[h_t]),[h])
     \frac{\delta\beta F^\text{ex}}{\delta \rho_\alpha(x,tH^t_\alpha(x,[h_t]))}[\set{\rho}^\text{eq}[h],
     \set{H}^+[h_+],\set{H}^-[h_-]]
   \label{eq:tot-var}\\
   &\ 
   + \frac{\delta\beta F^\text{ex}}{\delta J^t_\alpha(x)}[\set{\rho}^\text{eq}[h],\set{H}^+[h_+],
   \set{H}^-[h_-]]\Bigg) 
   + L_y \int\limits_{-H^-_\alpha(x,[h_-])}^{H^+_\alpha(x,[h_+])}\!\!\!\d z\;
     \rho_\alpha^\text{eq}(x,z,[h])\int\!\d x'\;\frac{\delta\beta V_\alpha}{\delta h_t(x')}(x,z,[h])
     \ \delta h_t(x')\Bigg],
   \notag
\end{align}
where we have substituted $J^\pm_\alpha(x)=H^\pm_\alpha(x,[h_\pm])$.
From Eq.~(\ref{eq:Fex}) one obtains
\begin{align}
   \frac{\delta\beta F^\text{ex}}{\delta\rho_\alpha(x,z)}\left[\set{\rho},\set{J}^+,\set{J}^-\right]
   &= -L_{y} \sum_{n=2}^\infty\frac{1}{(n-1)!} \sum_{\alpha_1,\dots,\alpha_{n-1}=1}^s 
   \int\!\d x_1\!\!\!\!\!\!\!\!\int\limits_{-J^-_{\alpha_1}(x_1)}^{J^+_{\alpha_1}(x_1)}\!\!\!\!\!\!\!\d z_1
   \dots
   \int\!\d x_{n-1}\!\!\!\!\!\!\!\!\int\limits_{-J^-_{\alpha_{n-1}}(x_{n-1})}^{J^+_{\alpha_{n-1}}(x_{n-1})}
   \!\!\!\!\!\!\!\d z_{n-1}\;
   \notag\\
   &\phantom{= -L_{y} \sum_{n=2}^\infty}
   c_{\alpha_1\dots\alpha_{n-1}\alpha}^{(0)}(x_1,z_1,\dots,x_{n-1},z_{n-1},x,z)
   \rho_{\alpha_1}(x_1,z_1)\cdots\rho_{\alpha_n}(x_{n-1},z_{n-1}).
   \label{eq:dFexdrho}
\end{align}
and
\begin{align}
   \frac{\delta\beta F^\text{ex}}{\delta J^\pm_\alpha(x)}\left[\set{\rho},\set{J}^+,\set{J}^-\right] =
   &\ -L_{y} \sum_{n=2}^\infty\frac{1}{(n-1)!} \sum_{\alpha_1,\dots,\alpha_{n-1}=1}^s 
   \int\!\d x_1\!\!\!\!\!\!\!\!\int\limits_{-J^-_{\alpha_1}(x_1)}^{J^+_{\alpha_1}(x_1)}\!\!\!\!\!\!\!\d z_1
   \dots
   \int\!\d x_{n-1}\!\!\!\!\!\!\!\!\int\limits_{-J^-_{\alpha_{n-1}}(x_{n-1})}^{J^+_{\alpha_{n-1}}(x_{n-1})}
   \!\!\!\!\!\!\!\d z_{n-1}\;
   \label{eq:dFexdJ}\\
   &\ 
   c_{\alpha_1\dots\alpha_{n-1}\alpha}^{(0)}(x_1,z_1,\dots,x_{n-1},z_{n-1},x,\pm J^\pm_\alpha(x))
   \rho_{\alpha_1}(x_1,z_1)\cdots\rho_{\alpha_n}(x_{n-1},z_{n-1})\rho_\alpha(x,\pm J^\pm_\alpha(x)),
   \notag
\end{align}
which leads to
\begin{align}
   -\rho_\alpha(x,\pm J^\pm_\alpha(x))
   \frac{\delta\beta F^\text{ex}}{\delta\rho_\alpha(x,\pm J^\pm_\alpha(x))}\left[\set{\rho},\set{J}^+,
   \set{J}^-\right]
   +
   \frac{\delta\beta F^\text{ex}}{\delta J^\pm_\alpha(x)}\left[\set{\rho},\set{J}^+,\set{J}^-\right] 
   = 0.
   \label{eq:remark}
\end{align}
Note that Eq.~(\ref{eq:remark}) is independent of the microscopic details of the interactions between the
fluids molecules. 
By substituting Eq.~(\ref{eq:remark}) into Eq.~(\ref{eq:tot-var}) one obtains the final expression for the
work necessary to deform the walls $h_\pm$ by $\delta h_\pm$: 
\begin{align}
   \delta\beta\Xi\left[h_\pm,\delta h\right] 
   &= L_y\int\!\d x \sum_{\alpha=1}^s\sum_{t\in\{\pm\}}\!\!\Bigg[
   -\rho_\alpha^\text{eq}(x,tH^t_\alpha(x,[h_t]),[h])^2f_\alpha'(\rho_\alpha^\text{eq}(x,tH^t_\alpha(x,
   [h_t]),[h]))\ \delta H^t_\alpha(x,[h_t,\delta h_t])
   \notag\\
   &\phantom{= L_y\int\!\d x \sum_{\alpha=1}^s\sum_{t\in\{\pm\}}\!\!\Bigg[}
   + \int\limits_{-H^-_\alpha(x,[h_-])}^{H^+_\alpha(x,[h_+])}\!\!\!\d z\;
      \rho_\alpha^\text{eq}(x,z,[h])\int\!\d x'\;\frac{\delta\beta V_\alpha}{\delta h_t(x')}(x,z,[h])
      \ \delta h_t(x')\Bigg].
   \label{eq:tot-var1}
\end{align}

In order to express the deformation work $\delta\beta\Xi\left[h,\delta h\right]$ in 
Eq.~(\ref{eq:tot-var1}) explicitly in terms of the variation $\delta h_\pm$ of the wall shapes $h_\pm$,
the boundaries $\set{H}^\pm[h_\pm]$ of the accessible volume have to be determined as functionals of $h_\pm$.
A particle of species $\alpha$ located at point $(x,y,\pm H^\pm_\alpha(x,[h_\pm]))$ touches the wall at
point $(\bar{x}^\pm_\alpha(x,[h_\pm]),y,\pm h_\pm(\bar{x}^\pm_\alpha(x,[h_\pm])))$ (see 
Fig.~\ref{fig:scheme}).
Since particles of species $\alpha$ have radius $R_\alpha$ and since wall and particles touch each other
tangentially, one can readily derive the following relations:
\begin{align}
   x = \bar{x}^\pm_\alpha(x,[h_\pm]) + 
        \frac{R_\alpha h_\pm'(\bar{x}^\pm_\alpha(x,[h_\pm]))}{\sqrt{1+h_\pm'(\bar{x}^\pm_\alpha(x,[h_\pm]))^2}}
   \quad\text{and}\quad
   H^\pm_\alpha(x,[h_\pm]) &= h_\pm(\bar{x}^\pm_\alpha(x,[h_\pm])) - 
        \frac{R_\alpha}{\sqrt{1+h_\pm'(\bar{x}^\pm_\alpha(x,[h_\pm]))^2}}.
   \label{eq:x2}
\end{align}
Upon variation with respect to $h_\pm$ in Eq.~(\ref{eq:x2}) leads to $\delta H^\pm_\alpha(x,[h_\pm,\delta 
h_\pm]) = \delta h_\pm(\bar{x}^\pm_\alpha(x,[h_\pm]))$, i.e., upon changing the wall shape the boundary of 
the accessible volume changes exactly as the wall shape does at the touching point.
Substituting the last expression into Eq.~(\ref{eq:tot-var1}) one obtains
\begin{align}
   \delta\beta\Xi\left[h,\delta h\right] 
   &= L_y\int\!\d x \sum_{\alpha=1}^s\sum_{t\in\{\pm\}}\!\!\Bigg[
   -\rho_\alpha^\text{eq}(x,tH^t_\alpha(x,[h_t]),[h])^2f_\alpha'(\rho_\alpha^\text{eq}(x,tH^t_\alpha(x,
   [h_t]),[h]))\ \delta h_t(\bar{x}^t_\alpha(x,[h_t]))
   \notag\\
   &\phantom{= L_y\int\!\d x \sum_{\alpha=1}^s\sum_{t\in\{\pm\}}\!\!\Bigg[}
   + \int\limits_{-H^-_\alpha(x,[h_-])}^{H^+_\alpha(x,[h_+])}\!\!\!\d z\;
      \rho_\alpha^\text{eq}(x,z,[h])\int\!\d x'\;\frac{\delta\beta V_\alpha}{\delta h_t(x')}(x,z,[h])\ 
      \delta h_t(x')\Bigg].
   \label{eq:tot-var2}
\end{align}
In the light of the first relation in Eq.~(\ref{eq:x2}), introducing the inverse 
maps of $x\mapsto\bar{x}^\pm_\alpha(x,[h_\pm])$,
\begin{align}
   K^\pm_\alpha(x',[h_\pm]) := x' + \frac{R_\alpha h_\pm'(x')}{\sqrt{1+h_\pm'(x')^2}},
   \label{eq:def-K}
\end{align}
one obtains 
\begin{align}
   \partial_{x'}K^\pm_\alpha(x',[h_\pm]) 
   = 1 + R_\alpha\frac{h_\pm''(x')}{\sqrt{1+h_\pm'(x')^2}^3}
   = 1 + \frac{R_\alpha}{R^\pm_\text{w}(x',[h_\pm])}.
\label{eq:k_alpha}
\end{align}
Then, finally, one can rewrite Eq.~(\ref{eq:tot-var2}) as Eq.~(\ref{eq:tot-var3}) in 
Sec.~\ref{subsec:workpressure} of the main text.

\end{widetext}


\section{Rotation invariance of $P_z$\label{app:Pz}}

One can show that $P^t_z$ is invariant upon rotation of the walls in the $x$-$z$ plane.
Indeed, consider the rotation $T(\varphi):\mathbb{R}^2\to\mathbb{R}^2, (x,z)\mapsto(\tilde{x},\tilde{z})$
around some center $(x_0,z_0)\in\mathbb{R}^2$ by an angle $\varphi$ given by
\begin{align}
   \left(\begin{array}{c}
      \tilde{x} \\
      \tilde{z}
   \end{array}\right)
   &=
   \left(\begin{array}{c}
      x_0 \\
      z_0
   \end{array}\right)
   +
   \mathbb{T}(\varphi)
   \left(\begin{array}{c}
      x-x_0 \\
      z-z_0
   \end{array}\right)
   \label{eq:rotation}
\end{align}
with the rotation matrix
\begin{align}
   \mathbb{T}(\varphi) =    
   \left(\begin{array}{cc}
      \cos\varphi & -\sin\varphi \\
      \sin\varphi &  \cos\varphi
   \end{array}\right)
   \in SO(2,\mathbb{R}).
   \label{eq:matrix}
\end{align}   

\begin{figure}[h!]
 \includegraphics[scale=0.35]{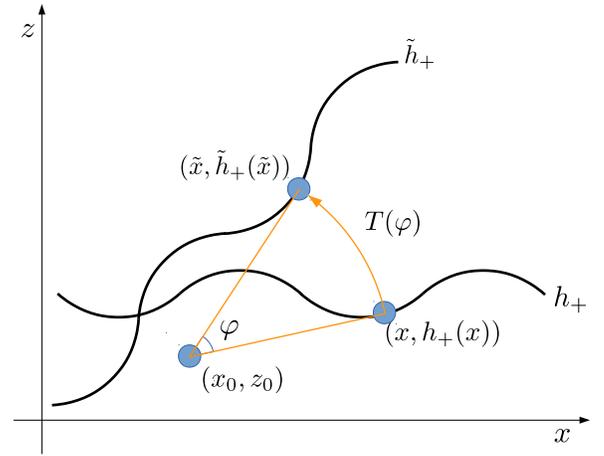}
 \caption{Cartoon of the tilted scenario}
 \label{fig:scheme2}
\end{figure}

Hence, the point $(x,th_t(x))$ at the wall of the original system is rotated by $T(\varphi)$ to the point
$(\tilde{x},t\tilde{h}_t(\tilde{x}))$ at the wall of the rotated system (see Fig.~\ref{fig:scheme2}).
As the two-dimensional Lebesgue-Borel measure $\d x\,\delta h_t(x)$ on the right-hand side of 
Eq.~(\ref{eq:tot-var4}) is invariant upon rotation, i.e., $\d x\,\delta h_t(x) = \d \tilde{x}\,\delta
\tilde{h}_t(\tilde{x})$, and as the work required to deform the walls is a scalar quantity, so is 
$P^t_z$:
\begin{align}
   P^t_z(\tilde{x},[\tilde{h}]) = P^t_z(x,[h]).
   \label{eq:rotinv}
\end{align}

\bibliography{nonequilibrium_pressure}

\end{document}